\documentclass[11pt,a4paper,dvipsnames]{article}
\usepackage{geometry}
\usepackage{mathrsfs}
\usepackage{microtype}
\usepackage[english]{babel}
\usepackage{multirow}
\usepackage{epigraph}
\usepackage{feynmp}
\usepackage{amsmath,amssymb}
\usepackage{hyperref}
\usepackage{float}
\usepackage{mathtools}
\usepackage{caption}
\usepackage{subcaption}
\usepackage{cite}
\usepackage{authblk}

\begin{document}
\title{\textbf{Holographic description of $F(R)$ gravity coupled with Axion Dark Matter}} 
\author[1]{Simone~D'Onofrio\,\thanks{donofrio@ice.csic.es}} \affil[1]{
 Institute of Space Sciences (ICE, CSIC) C. Can Magrans s/n, 08193 Barcelona, Spain.} 
 \date{}
\maketitle
\vspace{-1.cm}
\begin{abstract}
    In this work we study the autonomous dynamical system of different $F(R)$ models in the formalism of holographic dark energy using the generalized Nojiri-Odintsov cut-off. We explicitly give the expression of the fixed points as functions of the infrared cut-off for vacuum $F(R)$ gravity in flat and non-flat FRW background and for $F(R)$ coupling axion dark matter. Each fixed point component can be taken as a condition on the cut-off and on the expression of $F(R)$, leading to physically interesting constraints on these functions. 
\end{abstract}
\section{Introduction}

The importance of $F(R)$ modified gravity \cite{nojiri2017modified,capozziello2010beyond,nojiri2011unified} is given by the possibility to describe various eras of our Universe and the possibility to unify such scenarios in a single theory. The late time evolution is driven by dark energy and a description of such an era as a type of modified gravity can be found in \cite{capozziello2002curvature}. Several modified models provide a unification from an early-time acceleration to a late-time acceleration \cite{nojiri2003modified,nojiri2006modified,nojiri2011unified,capozziello2015connecting,cognola2008class,nojiri2007unifying,nojiri2008modified}, which provides a full description of the post-inflationary evolution. For reviews see \cite{faraoni2011landscape,nojiri2017modified,nojiri2007introduction,nojiri2011unified,olmo2011palatini}. 
Even if the behavior of dark matter has been deeply studied and described in different ways, its nature, as well as the interaction with other Standard model particles, is obscure. The latest attempts to describe such matter are related to axion physics \cite{odintsov2019f,cicoli2019geometrical,caputo2019radiative,fukunaga2019efficient} and could in principle give observable particles \cite{du2018search}. Among all the axion models, the most promising is the misalignment model \cite{marsh2016axion}, which is based on a broken Peccei-Quinn symmetry during inflation and all subsequent phases. Recent studies of such fields coupled with $F(R)$ gravity have given promising results in providing a consistent description \cite{Odintsov:2019evb,nojiri2020f,odintsov2020geometric,odintsov2020aspects}. \\
Holographic gravity is a theory that suggests that information in a volume of space can be encoded on its boundary. This theory is based on the holographic principle \cite{hooft1993dimensional,susskind1995world,witten1998anti,bousso2002holographic}, which states that the entropy (or information content) of a system is proportional to the area of its boundary, rather than its volume. This principle has been inspired by black hole thermodynamics and string theory and relates the infrared cut-off of a theory to the largest distance in a theory. A holographic description has been given both for the early-time era \cite{horvat2011holographic,paul2019holographic,nojiri2019holographic,elizalde2019viscous,bargach2020induced,nojiri2023holographic} and for the late-time scenario \cite{li2004model,wang2017holographic,elizalde2005dark,saridakis2008restoring,guberina2005hint,ito2005holographic,pavon2005holographic}. A unified description of such eras can be found in the recent paper \cite{nojiri2019holographic}, in which a covariant unified scenario is given. In this paper, as in most of the papers cited above, the choice of the infrared cutoff is taken to be the generalized Nojiri-Odintsov cutoff first introduced in \cite{nojiri2006unifying}, which is shown to reproduce a wide class of dark energy models \cite{nojiri2021different}. Another important achievement of the holographic description of cosmological fluids is the possibility to study different types of cosmological entropies that generalize the Bekenstein-Hawking one \cite{nojiri2022early,odintsov2023generalised}.
For the above reasoning is evident the importance to study the phase space of the cosmological dynamical system of the theory from the holographic point of view. The standard analysis of phase space in cosmological systems has been studied in depth in the past \cite{odintsov2017autonomous,boehmer2012jacobi,goheer2007dynamical,de2007phase,oikonomou2018autonomous,giacomini2017dynamical,odintsov2017phase,ivanov2012cosmological,shabani2013f,guo2013cosmological} (see \cite{bahamonde2018dynamical} for a review). In this paper we will provide the same formalism in the holographic picture for the $F(R)$ vacuum gravity in flat and non-flat FRW background and for the $F(R)$ gravity coupling with axion dark matter. We will also show a covariant method to compute the holographic cut-off for general $F(R)$ models first introduced in \cite{nojiri2017covariant}, which have then also been generalized to $F(R)$ coupled axion dark matter models \cite{Nojiri:2020wmh}. In the following work we will compute the infrared cut-off for these models in a spatially non-flat FRW background.\\
This work is organized as follows. In section \ref{sec:F(R)Axion} we will briefly review the analysis of $F(R)$ models coupled to axion dark matter in the hypothesis of non-flat background. We also provide a study of the late era scenario in order to understand the contribution of the curvature to the dark energy era. In section \ref{sec:Hol} we introduce the holographic formalism and we compute a possible cut-off in the case of non-flat background. In section \ref{sec:DynSys} we finally study the dynamical systems of the models studied in the previous part from a holographic point of view.

\section{$F(R)$ gravity with Axion Dark Matter Models}\label{sec:F(R)Axion}
The general $F(R)$ gravity in vacuum has an action
\begin{equation}\label{ActionNonInteracting}
    S = \frac{1}{2\kappa^2}\int d^4x \sqrt{-g} F(R) \ ,
\end{equation}
with $\kappa^2 = 8\pi G  =\frac{1}{M_P^2}$. By varying this action with respect to $g_{\mu\nu}$ we have the equations of motion to be 
\begin{equation}
    F'(R) R_{\mu\nu} - \frac{1}{2}F(R)g_{\mu\nu} - \nabla_\mu \nabla_\nu F'(R) + g_{\mu\nu} \Box F'(R)=0 \ ,
\end{equation}
where a prime indicates the partial derivative with respect to the curvature.
This action will be assumed to be immersed in a spatially curved FRW background 
\begin{equation}\label{FRWCurved}
    ds^2 = -dt^2 + a(t)^2 \left(\frac{dr^2}{1-Kr^2}+r^2 d\theta^2 + r^2 \sin\theta^2 d\phi^2 \right) \ ,
\end{equation}
where the constant $K$ is the curvature of the spacelike part of the metric (given by the hypersurface $t=\text{const}$), with $K=0,+1$  and $-1$, corresponding to flat, positive and negative curved spacetimes. For the following metric the Ricci scalar is
\begin{equation} \label{RicciScalar}
    R = 6\left(\Dot{H} + 2 H^2 + \frac{K}{a^2}\right) \ .
\end{equation}
With the following background metric the equations of motions take the form
\begin{equation}
    3 H^2 F'(R) =- \frac{F(R)}{2} + \frac{R \, F'(R)}{2} -3 H \Dot{F}'(R) -\frac{3 K F'(R)}{a^2}\ .
\end{equation}

The action of a modified $F(R)$ gravity non-minimally coupled to an axion scalar field $\phi$ is given by 
\begin{align}
    S =& \int d^4x \sqrt{-g} \frac{1}{2\kappa^2} F(R,\phi)  \\=& \int d^4x \sqrt{-g}\ \left( \frac{1}{2\kappa^2} F(R) +\frac{1}{2\kappa^2}  h(\phi) G(R)-\frac{1}{2}\partial_\mu \phi\partial^\mu\phi - V(\phi)  \right) \label{ActionInteracting} \ .
\end{align}
As done in the non-interacting case we will consider this action in a spatially curved FRW background with expression \eqref{FRWCurved}. In this case the first Friedmann equation for the interacting lagrangian is
\begin{align}\label{AxionFriedmannEq}
    3 H^2F'(R,\phi)=&- \frac{F(R,\phi)}{2} + \frac{R \, F'(R,\phi)}{2} -3 H \Dot{F}'(R,\phi) -\frac{3 K F'(R,\phi)}{a^2}+\kappa^2 \left( \frac{1}{2}\dot{\phi}^2 + V(\phi)  \right) \ ,
\end{align}
and the equation of motion of the axion field is obtained by variating the action with respect to this field 
\begin{equation}\label{EqMotAxion}
    \Ddot{\phi} + 3 H \dot{\phi} + V_\phi(\phi)-\frac{1}{2}F_\phi(R,\phi) = 0 \ ,
\end{equation}
where the subscript $\phi$ indicates a differentiation with respect to the field $\phi$. 
In order to obtain an expression of the unified action that is in agreement with all the steps of the universe evolution, we make the following assumptions.
The choice of $F(R)$ is
\begin{equation}
    F(R) = R + \frac{1}{36 H_i^2}R^2 \ ,
\end{equation}
which for small curvatures reproduces the Starobinsky inflation model that is known to be in agreement with the recent observational data, and for greater curvatures it describes the matter dominated epoch after inflation. 
Under this condition the gravitational field equation \eqref{AxionFriedmannEq} takes the expression
\begin{align}\label{AxionFriedmmannStarobinskyEq}
    &3 H^2\left( 1+\frac{R}{18 H_i^2}+h(\phi)G'(R)\right)=\frac{R^2}{72 H_i^2}- h(\phi)G(R)+\frac{1}{2}Rh(\phi)G'(R)-\frac{H\Dot{R}}{6 H_i^2}\nonumber \\ &-3H\Dot{R} h(\phi)G''(R) - \frac{3K}{a^2}\left( 1+\frac{R}{18 H_i^2}+h(\phi)G'(R)\right) +\kappa^2 \left( \frac{1}{2}\dot{\phi}^2 + V(\phi)  \right) \ .
\end{align}
For axion field in general we mean the pseudoscalar Goldstone boson of spontaneously symmetry breaking of the global chiral symmetry \cite{marsh2016axion} 
. In QCD the chiral symmetry is given by the Peccei Quinn rotation that acts on a field $\psi$ as 
\begin{equation}
    \psi \rightarrow e^{iQ\frac{\phi}{f_a}} \psi \ ,
\end{equation}
where $f_a$ is the axion decay constant and $Q$ the Peccei Quinn charge of the field $\phi$. At the classical level, the Lagrangian is invariant under this transformation, which means that it is invariant under the change of field $\phi \rightarrow \phi + \text{const}$. At quantum level the Peccei-Quinn rotations of quarks are anomalous and therefore this symmetry is broken at some scale $f_a$. The axion is the Goldstone boson of this symmetry breaking. At lower energy scales, $\Lambda_a < f_a$, non-perturbative effects switch on and break the shift symmetry, generating an induced potential $V(\phi)$. Since the axion is an angular degree of freedom of a complex scalar field, the relation $\phi\rightarrow \phi+ 2 n \pi f_a$ holds for some rational $n$. This symmetry tells that the potential $V(\phi)$ must be periodic in $\phi$ and therefore can be written as
\begin{equation}
    V(\phi) = \Lambda^4_a \, \, U\left(\frac{\phi}{f_a}\right) \ ,
\end{equation}
where $U$ is periodic. Because of the periodic condition the potential possesses a maximum and a minimum in the period. For small displacement from the minimum we can expand in power series the potential as
\begin{equation}
    V(\phi) \sim\frac{1}{2}m_a^2\phi^2 \ ,
\end{equation}
with $m_a = \frac{\Lambda_a^2}{f_a}$. The axion field is weakly interacting with itself and the other standard model's particles since the interaction is dumped by a factor of powers of $\frac{1}{f_a}$, as are the corrections to the axion mass given by quantum computations (due to the symmetry breaking). For these reasons the equation of motion of the axion field \eqref{EqMotAxion} will be entirely driven by the potential $V(\phi)$ and the following assumption will be taken
\begin{equation}
    V_\phi(\phi) \gg F_\phi(R,\phi) \ .
\end{equation}
Under this assumption the equation of motion is decoupled by the expression of $F(R,\phi)$ and, using the expansion of $V(\phi)$, it becomes
\begin{equation}
    \Ddot{\phi} + 3 H \dot{\phi} + m_a^2 \phi = 0 \ .
\end{equation}
As described before, the Peccei Quinn symmetry is broke in the pre-inflationary era and all the later scenario, resulting in the creation of a Goldstone boson, the axion field, which will have a non-zero vev. During the inflationary era ($H\gg m_a$) this field will not contribute, but in the later dynamic ($H\sim m_a$) the field will start to oscillate resulting in an effective energy density of the type of cold dark matter $\rho \sim a^{-3}$ for $H\ll m_a$ \cite{Odintsov:2019evb}.

\begin{figure}[htb]
     \centering
     \begin{subfigure}[b]{0.3\textwidth}
         \centering
         \includegraphics[width=\textwidth]{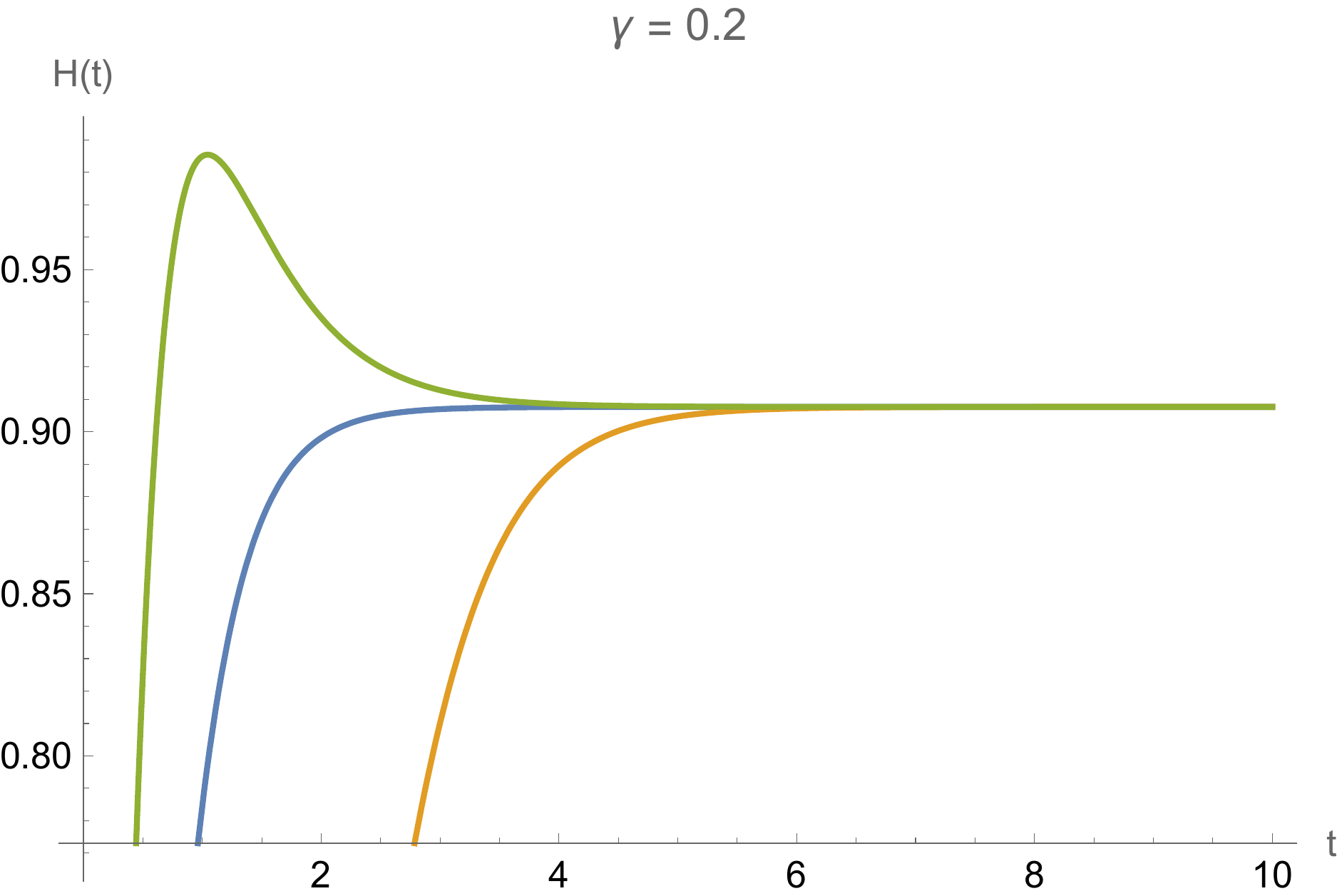}
     \end{subfigure}
     \hfill
     \begin{subfigure}[b]{0.3\textwidth}
         \centering
         \includegraphics[width=\textwidth]{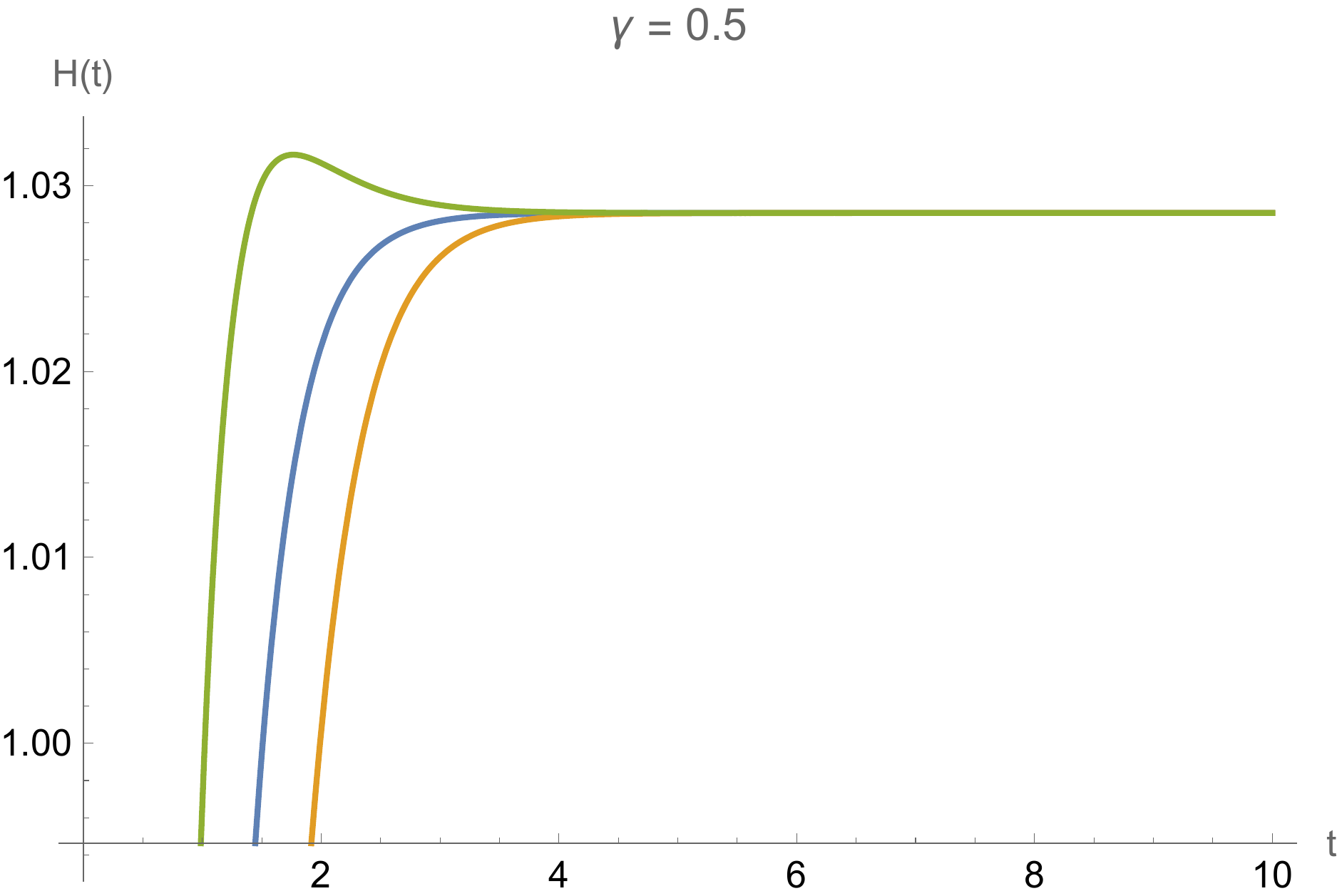}
     \end{subfigure}
     \hfill
     \begin{subfigure}[b]{0.3\textwidth}
         \centering
         \includegraphics[width=\textwidth]{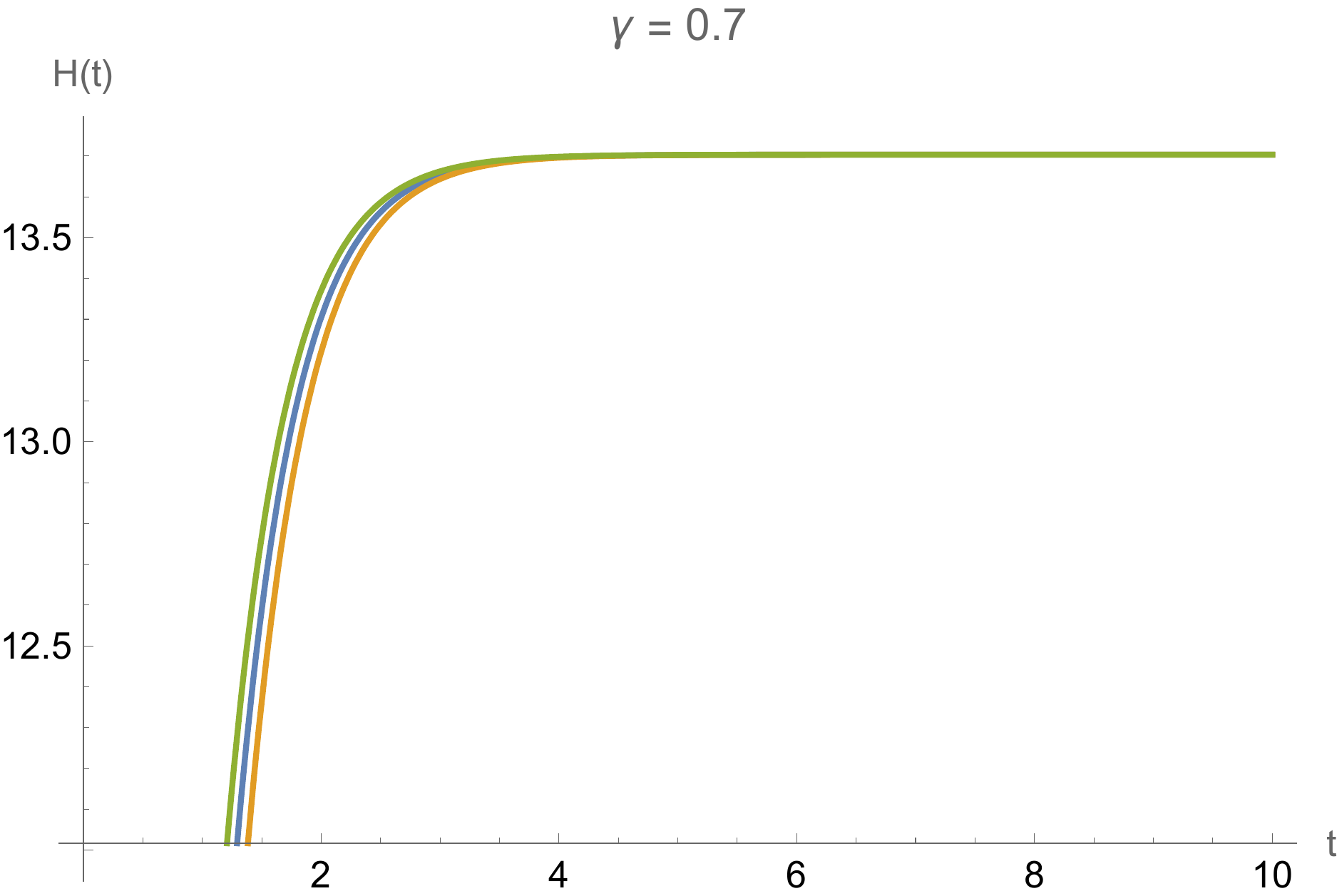}
     \end{subfigure}
        \caption{Numerical evaluation of the Hubble parameter given by the solution of the differential equation \eqref{DiffEqZ} for different values of $\gamma \in (0,\frac{3}{4})$. The blue line represent $K=0$, the orange line $K=1$ and the green line $K=-1$. The values of the constants are set to be $t_i=0$, $t_f = 10$, $\Lambda = 1$, $z(t_0) = 1^\frac{-1}{4\gamma-3}$ and $z'(t_0) = 0$.}
        \label{fig:NumericalH}
\end{figure}

We showed how the axion field dynamic gives a contribution to the universe evolution. Summarizing, in the early era this field is frozen at its vev and the inflationary dynamic is driven by gravity. Once the universe expands large enough that $H\sim m_a$, this field becomes relevant and gives a contribution to the effective energy density that mimics that of dark matter ($\rho_a \sim a^{-3}$). Before studying the fixed point of this model we shall analyze the asymptotic behavior of the cosmological solutions in the presence of non-null spatial curvature. We will see that this does not influence the asymptotic behavior of such model which is known to correctly reproduce dark matter behavior as said above. The inflationary scenario, large curvature regime, has been already studied in \cite{Odintsov:2019evb} and the presence of a non-null spatial curvature does not change the computations. In the small curvature regime, on the other hand, there is a contribution from this term and we will need to compute it.  In order to do so we will make use of the assumptions
\begin{equation}
    \kappa^2 V(\phi) \gg h(\phi)G(R) \ ,
\end{equation}
\begin{equation}
    h(\phi) \sim \frac{1}{\phi^\delta}
\end{equation}
with $\delta < 0$, 
and 
\begin{equation}
    G(R) \sim R^\gamma
\end{equation}
with $0<\gamma< \frac{3}{4}$. \\
Under this assumptions and for small curvature the dominant terms of \eqref{AxionFriedmmannStarobinskyEq} are given by
\begin{equation}
    3H^2h(\phi)G'(R) \simeq -3H\Dot{R}h(\phi)G''(R) \ ,
\end{equation}
which can be rewritten in 
\begin{equation}
    R H \simeq (1-\gamma)\Dot{R} \ .
\end{equation}
From the expression of the Ricci scalar \eqref{RicciScalar} and omitting subdominant terms of the form $H^3$ we get the differential equation
\begin{equation}
    (1-\gamma) \Ddot{H}+(3-4\gamma)H\Dot{H}-(3-2\gamma)K\frac{H}{a^2}=0 \ ,
\end{equation}
which can be rewritten as 
\begin{equation}
    \frac{d}{dt}\left(  (1-\gamma) \Dot{H}+\frac{3-4\gamma}{2}H^2+\frac{(3-2\gamma)K }{2 a^2}\right)=0 \ .
\end{equation}
Integrating this equation we have
\begin{equation}
    \Dot{H} - \beta H^2 +\alpha\frac{K}{a^2} = \frac{\Lambda_0}{1-\gamma} \ ,
\end{equation}
with $\Lambda_0$ integrating constant and $\beta=\frac{4\gamma-3}{2(1-\gamma)}$ and $\alpha =\frac{3-2\gamma}{2(1-\gamma)}$.
Using the definition of the Hubble parameter we can express the differential equation as a function of the scale parameter
\begin{equation}
    \frac{\Ddot{a}}{a}-(1+\beta)\left(\frac{\Dot{a}}{a}\right)^2  +\alpha\frac{K}{a^2} = \frac{\Lambda_0}{1-\gamma} \ .
\end{equation}
Thanks to the identity 
\begin{equation}
    \frac{d^2a^{-\beta}}{dt^2} = -\beta a^{-\beta} \left(\frac{\Ddot{a}}{a}-(1+\beta)\left(\frac{\Dot{a}}{a}\right)^2\right) 
\end{equation}
it can be simplified, applying the substitution $z(t) = a(t)^{-\beta}$, in the following equation
\begin{align}
     \frac{d^2z(t)}{dt^2} =& -\frac{\Lambda_0\beta}{1-\gamma} z(t) + K \alpha\beta z(t)^\frac{\beta+2}{\beta} \label{DiffEqZBeta} \\ 
     = &\Lambda_0\frac{3-4\gamma}{2(\gamma-1)^2} z(t) + K\frac{(4\gamma-3)(3-2\gamma)}{4(\gamma-1)^2}z(t)^\frac{1}{4\gamma-3}\label{DiffEqZ}  \ ,
\end{align}
where in the last identity we substituted back $\alpha$ and $\beta$. \\
In the case $K=0$ we find the solution to be
\begin{equation}
    z(t) \sim \cosh{\left(\sqrt{\frac{\Lambda_0\beta}{\gamma-1}}(t-t_0)\right)} \ ,
\end{equation}
and therefore the scale parameter is
\begin{equation}
    a(t) \sim \cosh{\left(\sqrt{\frac{\Lambda_0\beta}{\gamma-1}}(t-t_0)\right)}^{-\frac{1}{\beta}} \ ,
\end{equation}
with a resulting Hubble parameter of the form
\begin{align}
    H(t) \sim & \sqrt{\frac{\Lambda_0}{\beta(\gamma-1)}}\tanh{\left(\sqrt{\frac{\Lambda_0\beta}{\gamma-1}}(t-t_0)\right)}\\=& \sqrt{\frac{2\Lambda_0}{3-4\gamma}}\tanh{\left(\sqrt{\frac{\Lambda_0(3-4\gamma)}{2(\gamma-1)^2}}(t-t_0)\right)}\ ,
\end{align}
where in the last equation we have substitute back $\beta$ in order to have a clean expression of $H$.
This expression is in perfect agreement with the one calculated in \cite{Odintsov:2019evb}, with the only difference in the choice of constant $\Lambda = \frac{\Lambda_0}{1-\gamma}$.
\\
It is not possible to analytically solve the differential equation \eqref{DiffEqZ} for general values of $\gamma$ and fixing $K=\pm 1$. But it seems possible to numerically solve this equation with reasonable boundary conditions and under the hypothesis of $0<\gamma<\frac{3}{4}$. So we have computed the solution $z(t)$ to the ODE \eqref{DiffEqZ} for three different values of $\gamma$ and each for all the values of $K$. Then we have computed the Hubble parameter as a function of $z(t)$ as $H(t) = -\beta\frac{\Dot{z}}{z}$ and plotted it in the integration interval. The result of this analysis is given in figure \ref{fig:NumericalH} with the numerical values of the integrating constants. We see that the effects of spatial curvature have no implication of the late time dynamics as expected and also that the effects on this term on the Hubble parameter are inversely proportional to $\gamma$.

\section{Holographic description of gravity in spatially curved space-time}\label{sec:Hol}
In the following section we will follow the procedure given in \cite{Nojiri:2020wmh,nojiri2017covariant} to build an infrared cut-off which reproduces the gravitational field equations in the hypothesis of a spatially curved FRW background. This choice of cut-off takes the name of Nojiri-Odintsov cut-off. We can express the holographic dark energy density as a function of the infrared cut-off $L_{IR}$ as
\begin{equation}\label{rhoHol}
    \rho_{\text{hol}} = \frac{3 c^2}{\kappa^2 L_{IR}^2} \ .
\end{equation}
In a flat FRW background the Friedmann equation reads
\begin{equation}
    H^2=\frac{\kappa^2}{3}\rho_{\text{hol}}
\end{equation}
and using \eqref{rhoHol} we rewrite the Friedmann equation in the holographic formalism as
\begin{equation}\label{holEq}
    H = \frac{c}{L_{IR}} \ .
\end{equation}
The holographic correspondence states that for any expression of $F(R)$ we can find a $L_{IR}$ that, along with the holographic equation \eqref{holEq}, can reproduce the gravitational field equations of the modified gravity, both in vacuum that in non-vacuum models.
We have different possibilities for $L_{IR}$. We can choose it to be the particle horizon
\begin{equation}\label{Lp}
    L_{IR} = L_p = a \int^t_0 \frac{dt'}{a} 
\end{equation}
or the future event horizon 
\begin{equation}\label{Lf}
    L_{IR} = L_f = a \int^\infty_t \frac{dt'}{a} \ .
\end{equation}
With this cut-off can be prove that the theory leads to a de Sitter solution, which is known to not properly describe the inflationary scenario. \\
We may consider a more general cut-off which can be a function of $L_p$ and $L_f$ and their derivatives. Moreover, this cut-off can depend in general on the scale factor and to the Hubble parameter and its derivatives. This summarizes in setting 
\begin{equation}\label{L=L(Lp,Lp'..)}
    L_{IR} = L_{IR} (L_p,\Dot{L}_p, \dots,L_f,\Dot{L}_f, \dots, a, H, \Dot{H}, \dots) \ .
\end{equation}
The above cut-off is called generalized Nojiri-Odintsov cut-off and was firstly introduced in \cite{nojiri2006unifying} and allow us to describe any dark energy model \cite{nojiri2021different}. In \cite{Nojiri:2020wmh,nojiri2017covariant} a method is given to compute this generalized cut-off for a general $F(R)$ theory in the FRW flat background. In this section we will derive this cut-off in a generally spatially curved FRW background \eqref{FRWCurved} both for the only $F(R)$ gravity and for the axion coupled $F(R)$ gravity. \\ 
For the non-interacting case the action \eqref{ActionNonInteracting} will be rewritten as 
\begin{equation}
    S = \frac{1}{2\kappa^2}\int d^4x \sqrt{-g} \, F(R)  = \frac{1}{2\kappa^2}\int d^4x \sqrt{-g} \big(R+f(R)\big) \ .
\end{equation}
Under this assumption the gravitational field equation are given by
\begin{equation}
    3 H^2 = -\frac{f(R)}{2}+3(H^2+\dot{H}) f'(R) - 3 H \Dot{f'}(R) - \frac{3 K}{a^2} \ ,
\end{equation}
identifying the Hubble parameter in the left side of the equation with the holographic cut-off, thanks to equation \eqref{holEq}, we obtain
\begin{equation}
    \frac{3 c^2}{\left(L_{IR}\right)^2} = -\frac{f(R)}{2}+3(H^2+\dot{H}) f'(R) - 3 H \Dot{R}f''(R) - \frac{3 K}{a^2} \ .
\end{equation}
The right hand side of this equation can be written as a function of $L_p$ (or $L_f$), the derivatives and the scale parameter $a$, providing a cut-off of the type \eqref{L=L(Lp,Lp'..)}. In order to do so we rewrite the Hubble parameter in terms of the horizons as
\begin{align}
    H&= \frac{\Dot{L}_p}{L_p}-\frac{1}{L_p} \label{H_Lp}\\
    &=\frac{\Dot{L}_f}{L_f}+\frac{1}{L_f} \label{H_Lf} \ ,
\end{align}
which provides, using the equation \eqref{RicciScalar}, a Ricci scalar of the type
\begin{align}
    R &= 6\left( \frac{\Ddot{L}_p}{L_p} +\frac{\Dot{L}_p^2}{L_p^2}-\frac{3\Dot{L}_p}{L_p^2}+\frac{2}{L_p^2} + \frac{K}{a^2}\right)  \label{R_Lp}\\
     &= 6\left( \frac{\Ddot{L}_f}{L_f} +\frac{\Dot{L}_f^2}{L_f^2}+\frac{3\Dot{L}_f}{L_f^2}+\frac{2}{L_f^2}+ \frac{K}{a^2}\right) \label{R_Lf}\ .
\end{align}
Using these relations we can rewrite the cut-off as 
\begin{align}
\frac{3c^2}{L_{IR}^2} =& -\frac{f(R_{(L_p)})}{2} +3\left( \frac{\Ddot{L}_p}{L_p}-\frac{\Dot{L}_p}{L_p^2}+\frac{1}{L_p^2}\right)f'(R_{(L_p)})- 3 \left(\frac{\Dot{L}_p}{L_p}-\frac{1}{L_p}\right) \Dot{R}_{(L_p)} f''(R_{(L_p)})  - \frac{3 K}{a^2}\label{CutOffp}\\
=& -\frac{f(R_{(L_f)})}{2} +3\left( \frac{\Ddot{L}_f}{L_f}+\frac{\Dot{L}_f}{L_f^2}+\frac{1}{L_f^2}\right)f'(R_{(L_f)})- 3 \left(\frac{\Dot{L}_f}{L_f}+\frac{1}{L_f}\right) \Dot{R}_{(L_f)} f''(R_{(L_f)})  - \frac{3 K}{a^2}\ , \label{CutOfff}
\end{align}
which are two of the many generalized cut-off that, along with the equation \eqref{holEq} (and the expression of the Ricci scalar \eqref{R_Lp} and \eqref{R_Lf}), provides the gravitational equation for the $F(R)$ gravity in the FRW curved space-time. From \eqref{R_Lp} an \eqref{R_Lf} we obtain the expressions of 
\begin{align}
    \Dot{R} &= 6 \left(\frac{\dddot{L}_p}{L_p}-\frac{3\Ddot{L}_p}{L_p^2}+\frac{\dot{L}_p\Ddot{L}_p}{L_p^2}-\frac{2\dot{L}_p^3}{L_p^3}+\frac{6\dot{L}_p^2}{L_p^3}-\frac{4\dot{L}_p}{L_p^3}\right) \\
    &= 6 \left(\frac{\dddot{L}_f}{L_f}+\frac{3\Ddot{L}_f}{L_f^2}+\frac{\dot{L}_f\Ddot{L}_f}{L_f^2}-\frac{2\dot{L}_f^3}{L_f^3}-\frac{6\dot{L}_f^2}{L_f^3}-\frac{4\dot{L}_f}{L_f^3}\right)\ .
\end{align}

In the presence of matter fields, described by adding a term $\mathcal{L}_{\text{mat}}$ to the action \eqref{ActionNonInteracting}, the holographic correspondence can be obtained with the same cut-off provided in the vacuum case \eqref{CutOffp} and \eqref{CutOfff}, but with a gravitational equation of the type
\begin{equation}\label{holEqMatter}
    3H^2 = \frac{3 c^2}{L_{IR}^2} + \kappa^2\rho_{\text{mat}} \ .
\end{equation}
In the case of an $F(R)$ gravity non-minimally coupled with an axion field \cite{Nojiri:2020wmh} the action \eqref{ActionInteracting} will be rewritten in the following way
\begin{equation}
    S = \int d^4x \sqrt{-g}\ \left( \frac{1}{2\kappa^2}\big(R + f(R,\phi)\big)-\frac{1}{2}\partial_\mu \phi\partial^\mu\phi - V(\phi)  \right) \ ,
\end{equation}
with $f(R,\phi)=f(R)+  h(\phi) G(R)$. The cut-off expression will have the same form of the equations \eqref{CutOffp} and \eqref{CutOfff} with the difference that now the function $f$ also depends on the axion field $\phi$. Assuming the variation of this field with respect to cosmic time negligible compared to the variation of the curvature we can write $\Dot{f'}(R,\phi) \sim \Dot{R}f''(R,\phi)$, and therefore $L_{IR}$ is given by
\begin{align}
\frac{3c^2}{L_{IR}^2} =& -\frac{f(R_{(L_p)},\phi)}{2} +3\left( \frac{\Ddot{L}_p}{L_p}-\frac{\Dot{L}_p}{L_p^2}+\frac{1}{L_p^2}\right)f'(R_{(L_p)},\phi)\nonumber\\&- 3 \left(\frac{\Dot{L}_p}{L_p}-\frac{1}{L_p}\right) \Dot{R}_{(L_p)} f''(R_{(L_p)},\phi)  - \frac{3 K}{a^2}\\
=& -\frac{f(R_{(L_f)},\phi)}{2} +3\left( \frac{\Ddot{L}_f}{L_f}+\frac{\Dot{L}_f}{L_f^2}+\frac{1}{L_f^2}\right)f'(R_{(L_f)},\phi)- 3 \left(\frac{\Dot{L}_f}{L_f}+\frac{1}{L_f}\right) \Dot{R}_{(L_f)} f''(R_{(L_f)},\phi)  - \frac{3 K}{a^2}\ ,
\end{align}
which provides the needed cut-off which, along with the holographic equation \eqref{holEq}, reproduces the cosmological equation of the $F(R)$ gravity coupled axion dark matter in the non-flat FRW background.

\section{Dynamical system}\label{sec:DynSys}

In this section we will derive the autonomous dynamical system for the different models of modified gravity seen in the previous sections. Each dynamical system will be then translated and analyzed in the holographic formalism.\\
In the case of the vacuum $F(R)$ gravity in a flat FRW background the equations of motion are
\begin{align}
    3 H^2 F' &= -\frac{F}{2}+\frac{R}{2}F' - 3 H \Dot{F'} \\
    \Ddot{F'} &= H \Dot{F'}-2 \dot{H}F' \ ,
\end{align}
with $R=6(\Dot{H}+2H^2)$ . The first equation suggests the change of variables 
\begin{equation}\label{DynVarFlat}
  \hfill  x_1 = -\frac{\dot{F'}}{H F'} \hspace{1cm} x_2 = -\frac{F}{6H^2 F'} \hspace{1cm} x_3 = \frac{R}{6H^2} \hfill \ ,
    \end{equation}
under which it becomes $x_1+x_2+x_3=1$. Thanks to this constraint and using the relations
\begin{align}
    &\frac{\Dot{H}}{H^2} = x_3-2 \\
    &\frac{\Dot{R}}{H^3} = 6 ( 4(x_2-2)-m ) \\
    &\frac{\Ddot{F'}}{H^2 F'} =-x_1-2(x_3-2) \ ,
\end{align}
we see that this cosmological model can be rewritten, using the derivative with respect to the e-fold number defined as $\frac{d}{dN} = \frac{1}{H}\frac{d}{dt}$, as 
\begin{equation}\label{DynSystFlat}
\begin{aligned}
    \frac{d x_1}{dN} &= -4+3x_1+2x_3-x_1x_3+x_1^2 \\
     \frac{d x_2}{dN} &= 8 + m -4 x_3 + x_2x_1-2x_2x_3+4x_2\\
      \frac{d x_3}{dN} &=-8-m+8x_3-2x_3^2 \ ,
\end{aligned}
\end{equation}
with $ m \equiv -\frac{\Ddot{H}}{H^3}$. The explicit dependence on cosmic time of this system of differential equations is contained only in the term $ m$. In the hypothesis of $ m = \text{const}$ the system becomes autonomous and we can analyze the attractors in the phase space to describe the stable asymptotic solutions of these cosmologies and compute which $F(R)$ theory realizes such condition. The condition $m \simeq 0$ is interesting since it is equivalent to the slow-roll condition, and the study of the attractors under this condition gives information about the inflationary era, the respective fixed point are de Sitter attractors. In the case $m = -\frac{9}{2}$ we would get the matter dominated fixed points ($w_\text{eff} = 0$), and for $m=-8$ the radiation domination fixed points ($w_\text{eff}=\frac{1}{3}$).  The EoS of such model is described by the parameter 
\begin{equation}\label{weff}
    w_{\text{eff}} = -1-\frac{2\Dot{H}}{H^2} = \frac{1}{3}-\frac{2}{3}x_3 \ .
\end{equation}\\
This formalism can be rewritten in the holographic picture using the holographic equation $H = \frac{c}{L_{IR}}$ which relates the Hubble parameter to the infrared cut-off. In a flat FRW background the scalar curvature can be expressed as a function of the cut-off and its derivative as
\begin{equation}\label{CurvatureCutOff}
    R_{(L_{IR})} = \frac{6c}{L_{IR}^2}\left(2c-\Dot{L}_{IR}\right) \ .
\end{equation}
The variables \eqref{DynVarFlat} become then
\begin{equation}\label{DynVarFlatHol}
  \hfill  x_1 = -\frac{L_{IR} \, \dot{F'}}{c \, F'} \hspace{1cm} x_2 = -\frac{L_{IR}^2 \, F}{6 \, c^2 \,  F'} \hspace{1cm} x_3 = 2- \frac{\Dot{L}_{IR}}{c} \hfill \ ,
    \end{equation}
where the function $F(R)$ has to be considered as a function of \eqref{CurvatureCutOff}. The dynamical system \eqref{DynSystFlat} still holds for the new variables with the substitution
\begin{equation}
    m = \frac{\Ddot{L}_{IR}L_{IR}}{c^2}- 2(2-x_3)^2 \ .
\end{equation}
As said before in the hypothesis of $m=\text{const}$ the system becomes autonomous, this condition translates in the holographic framework in the condition 
\begin{equation}\label{mHolConst}
\Ddot{L}_{IR}L_{IR}-2\Dot{L}^2_{IR}=\text{const}
\end{equation}
Since the dynamical system is the same all the analysis in phase space (the fixed points and their stability) holds true for the holographic description with the new variables \eqref{DynVarFlatHol}.
The EoS parameter is obtained by \eqref{weff} as 
\begin{equation}
    w_{\text{eff}} = -1 +\frac{2}{3}\frac{\dot{L}_{IR}}{c} \ .
\end{equation} 
As shown in \cite{odintsov2017autonomous} the fixed points of the dynamical system \eqref{DynSystFlat} in the hypothesis of $m \simeq 0$ are 
\begin{align}
    (x_1,x_2,x_3)=\phi_1^* &= (-1,0,2) \hspace{0.7cm} \text{stable}\\ 
    \phi_2^* &=(0,-1,2)\hspace{0.7cm} \text{unstable} \ ,
    \end{align}
    both of them corresponds to a $w_\text{eff} = -1$ and so both of them are de Sitter attractors.
    The condition $m \simeq 0 $ in the holographic description fixes the expression of $L_{IR}$ to be 
    \begin{equation}\label{cutOffmO}
        L_{IR} = \frac{c}{H_0-H_i t} \ .
    \end{equation}
From these fixed points is possible to reconstruct the equivalent $F(R)$ which correspond to each of this attractors
\begin{align}\label{FixedF(R)}
    \phi_1^* &\rightarrow F(R) \simeq\Lambda_1 - 24 \Lambda_2 e^{-\frac{R}{24H_i}}  &&\text{stable}\\ 
    \phi_2^* &\rightarrow F(R) \simeq  \alpha R ^2 &&\text{unstable} \ ,
\end{align}
where again $R$ has to be meant as $R_{(L_{IR})}$ \eqref{CurvatureCutOff}. From this analysis we see that the Starobinsky $R^2$ gravity shows an unstable behavior which is good since it indicates an exit from the inflationary scenario, while the stable attractor is described by a cosmological constant plus exponential $F(R)$ gravity, which describes a realistic evolution after the inflationary era. We note that for the point $\phi_1^*$ we first have to impose the condition on the variable $x_1$ that fixes the expression of the $F(R)$ as \eqref{FixedF(R)} and then using $x_2$ to fix a condition on the parameters appearing in this function as done \cite{odintsov2017autonomous}. If we had done it the other way around, imposing first the condition on $x_2$, we would have obtained a null $F(R)$ which has no physical meaning.

For the vacuum $F(R)$ model in a curved FRW background  the dynamical system acquires a new degree of freedom due to the mean curvature of the space-time. The equations of motion read
\begin{align}
    3 H^2 F' &= -\frac{F}{2}+\frac{R}{2}F' - 3 H \Dot{F'} - \frac{3 K F'}{a^2}\\
    \Ddot{F'} &= H \Dot{F'}-2\dot{H}F' +\frac{2 K F'}{a^2}\ ,
\end{align}
with $ R = 6\left(\dot{H}+2H^2 + \frac{K}{a^2}\right)$. 
From the first equation we define the new variables
\begin{equation}\label{DynVarCurv}
  \hfill  x_1 = -\frac{\dot{F'}}{H F'} \hspace{1cm} x_2 = -\frac{F}{6H^2 F'} \hspace{1cm} x_3 = \frac{R}{6H^2}\hspace{1cm} x_4 = -\frac{K}{a^2H^2} \hfill \ ,
\end{equation}
which differ from the ones in the flat case by the addition of the new variable $x_4$ describing the mean curvature of the space-time. Using the identities 
\begin{align}
&1 = x_1+x_2+x_3+x_4 \\ 
   & \frac{\Dot{H}}{H^2} = x_3+x_4-2 \\
    &\frac{\Dot{R}}{H^3} = 6 (4x_3+6x_4-8-m) \\
    &\frac{\Ddot{F'}}{H^2 F'} =4-x_1-2x_3-4x_4\ ,
\end{align}
we obtain the dynamical system 
\begin{equation}\label{DynSystCurv}
\begin{aligned}
    \frac{d x_1}{dN} &=-4+3x_1+2x_3-x_1x_3+x_1^2+4x_4-x_1x_4 \\
     \frac{d x_2}{dN} &=8+m-4x_3+x_1x_2-2x_2x_3+4x_2-6x_4-2x_2x_4 \\
      \frac{d x_3}{dN} &= -8-m+8x_3-2x_3^2+6x_4-2x_3x_4 \\ 
      \frac{d x_4}{dN} &=-2x_3x_4-2x_4^2+4x_4-2x_4\ . 
\end{aligned}
\end{equation}
In this case the EoS parameter \eqref{weff} depends also on the variable $x_4$ and it has the expression
\begin{equation} \label{weffCurv}
    w_{\text{eff}} =-1-\frac{2}{3}\left(x_3+x_4-2\right) \ .
\end{equation}
We can rewrite the above model using the holographic equation \eqref{holEq} as described for the flat case. The only difference is the new variable $x_4$ which depends on the scale factor $a(t)$. We can express this factor in terms of the cut-off as
\begin{equation}\label{scaleParameterHol}
    a(t) = a(t_0) e^{\int_{t_0}^t\frac{dt'}{L_{IR}}} \ ,
\end{equation}
and therefore, taking  $a(t_0)=1$, we find the curvature to be
\begin{equation}
    R_{(L_{IR})} = 6 \left( \frac{2c^2}{L_{IR}^2}-\frac{c\dot{L}_{IR}}{L_{IR}^2}+Ke^{-2\int_{t_0}^t\frac{dt'}{L_{IR}}} \right)
\end{equation}
The variables \eqref{DynVarCurv} can be so rewritten in terms of the holographic cut-off as
\begin{equation}\label{DynVarCurvHol}
    \begin{aligned}
  \hfill  x_1 = -\frac{L_{IR} \, \dot{F'}}{c \, F'} \hspace{0.7cm} x_2 = -&\frac{L_{IR}^2 \, F}{6 \, c^2 \,  F'} \hspace{0.7cm} x_3 = 2- \frac{\Dot{L}_{IR}}{c}+\frac{K L_{IR}^2}{c^2}e^{-2\int_{t_0}^t\frac{dt'}{L_{IR}}}\hspace{0.7cm}\\ &\hfill x_4=-\frac{K L_{IR}^2}{c^2}e^{-2\int_{t_0}^t\frac{dt'}{L_{IR}}} \hfill \ .
\end{aligned}
\end{equation}

 The dynamical system \eqref{DynSystCurv} still holds for the new variables with the substitution
\begin{equation}\label{mCurved}
    m = \frac{\Ddot{L}_{IR}L_{IR}}{c^2}- 2(2-x_3-x_4)^2 \ , 
\end{equation}
and again can me rendered autonomous under the condition \eqref{mHolConst}. Since the dynamical system is the same, all the phase space analysis in the holographic framework is analogous, as it should, to the standard treatment. Therefore, all fixed points and their stability are the same with the only difference that the variables now are expressed in terms of the holographic cutoff \eqref{DynVarCurv}. The fixed points in the hypothesis of $m \simeq 0$ are  \cite{odintsov2019effects} 
\begin{align}
    (x_1,x_2,x_3,x_4)=\phi_1^* &= (-1,0,2,0) \hspace{0.7cm} \text{unstable}\\ 
    \phi_2^* &=(0,-1,2,0)\hspace{0.7cm} \text{unstable} \ ,
    \end{align}
    where from a first look seems that the presence of the spatial curvature does not affect the fixed point since for both $x_4 = 0$ and the leading form of $F(R)$ is still \eqref{FixedF(R)}, but in this case there are no stable fixed points for the system. Also, for other values of $m$ the non-flat analysis leads to unstable fixed points. So the existence of a non-null spatial curvature destabilizes the attractors of the cosmological model.

\begin{figure}[H]
    \centering
    \includegraphics[scale=0.37]{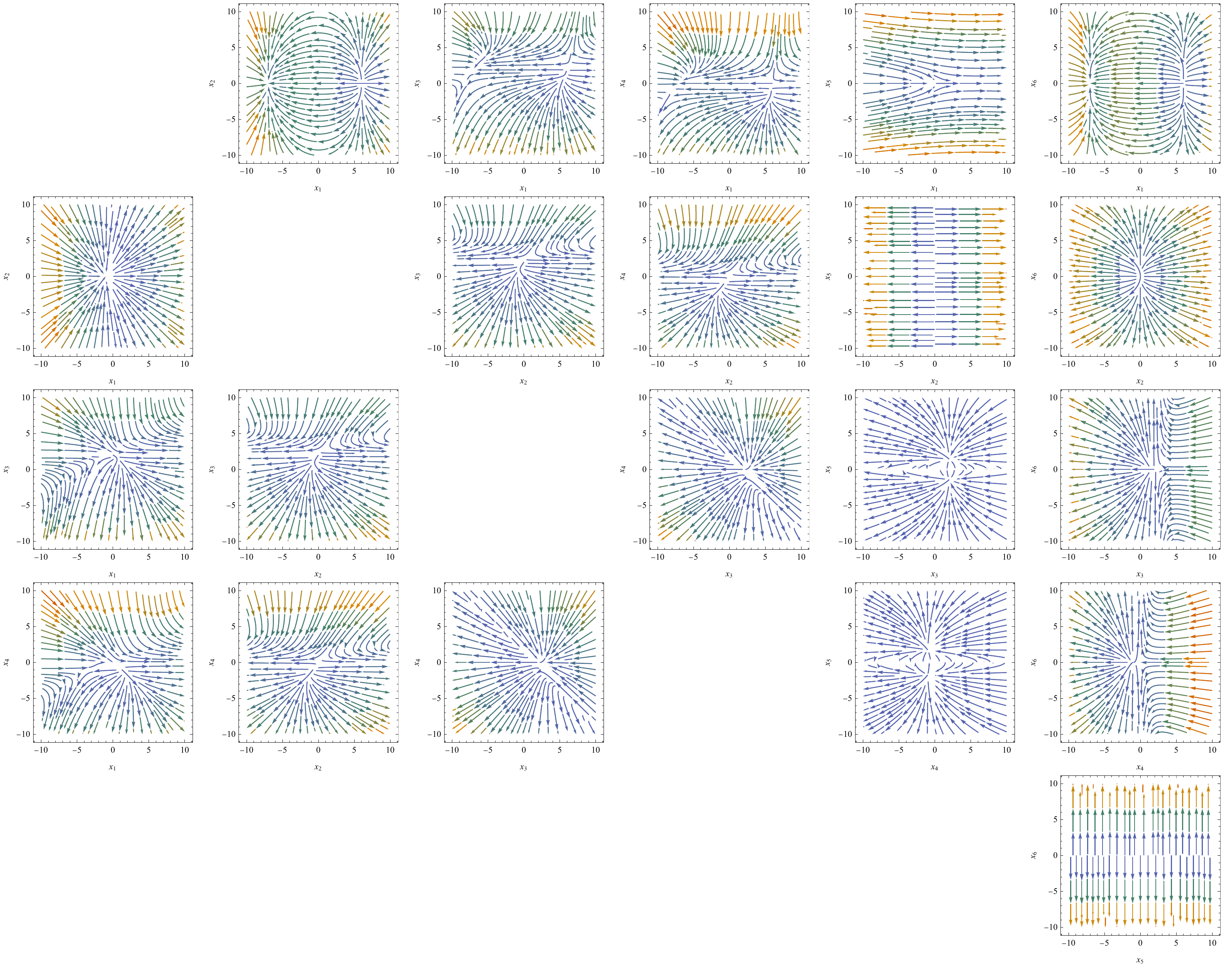}
    \caption{Trajectories in the $x_i-x_j$ planes for the dynamical systems \eqref{DynSystCurv} and \eqref{DynSystCurvAxion}. The plots under the diagonal refer to the vacuum $F(R)$ gravity in non-flat background \eqref{DynSystCurv}, the plots above the diagonal refer to the $F(R)$ gravity coupling with Axion Dark Matter in non-flat FRW background \eqref{DynSystCurvAxion}. The dynamics have been computed for $m=0$ and $\Gamma=0$.}
    \label{fig:phaseSpace}
\end{figure}

We will now derive the autonomous dynamical system for a $F(R)$ gravity coupled with an axion field $\phi$. The dynamical system for the General Relativity coupled Axion Dark Matter case has been computed firstly in \cite{copeland1998exponential} (also \cite{bahamonde2018dynamical} for a review). The general case for $F(R)$ gravity can be obtained using the same reasoning as we will show in the following.
The action is
\begin{equation}
        S = \int d^4x \sqrt{-g}\ \left( \frac{1}{2\kappa^2} F(R,\phi)-\frac{1}{2}\partial_\mu \phi\partial^\mu\phi - V(\phi)  \right) \ .
\end{equation}
A differentiation with respect to the metric gives the Friedmann equations, which in a curved FRW background take the expressions
\begin{align}
    3 H^2 F' &= -\frac{F}{2}+\frac{R}{2}F' - 3 H \Dot{F'} - \frac{3 K F'}{a^2} + \kappa^2\left(\frac{1}{2}\dot{\phi}^2+V \right)\\
     \Ddot{F'} &= H \Dot{F'}-2\dot{H}F' +\frac{2 K F'}{a^2}-\kappa^2\dot{\phi}^2 \ ,
\end{align}
along with the continuity equation
\begin{equation}
    \Ddot{\phi} + 3H\dot{\phi} + V_\phi -\frac{1}{2} F_\phi =0  \ .
\end{equation}
As firstly noted by \cite{copeland1998exponential} for the GR case ($F(R) = R$) we can render autonomous this dynamical system for a suitable choice of potentials $V(\phi)$. We can generalize this treatment for the $F(R)$ case, more precisely we make the following choice of variables   
\begin{equation}\label{DynVarCurvAxion}
\begin{aligned}
  \hfill  x_1 = -\frac{\dot{F'}}{H F'} \hspace{1cm} x_2 = -\frac{F}{6H^2 F'} \hspace{1cm} x_3 = \frac{R}{6H^2}\hspace{1cm} x_4 = -\frac{K}{a^2H^2} \hfill  \\
   \hfill  x_5 = \frac{\kappa\dot{\phi}}{\sqrt{6}H\sqrt{F'}} \hspace{1cm} x_6 =\frac{\kappa\sqrt{V}}{\sqrt{3}H\sqrt{F'}} \hspace{1cm} x_7 = \sqrt{F'}\frac{V_\phi}{\kappa V}\ . \hspace{1cm}
\end{aligned}
\end{equation}
The variable $x_7$ doesn't explicitly appears in the Friedmann constraint but it is needed to delete the explicit time dependence of the dynamical system. We will see that its expression is such that its equation of motion contains only a time dependent parameter, which can be taken constant under some reasonable conditions on the potential $V(\phi)$. Under this hypothesis then we can render autonomous the dynamical system. In order to compute this system we use the identities
\begin{align}
&1 = x_1+x_2+x_3+x_4 +x_5^2+x_6^2\\ 
   & \frac{\Dot{H}}{H^2} = x_3+x_4-2 \\
    &\frac{\Dot{R}}{H^3} = 6 (4x_3+6x_4-8-m) \\
    &\frac{\Ddot{F'}}{H^2 F'} =4-x_1-2x_3-4x_4-6x_5^2 \\
    &\frac{\Ddot{\phi}}{H^2\sqrt{F'}} = -3\sqrt{6}x_5 + \frac{\sqrt{6}}{2x_5}(4x_3+6x_4-8-m)-3x_6^2x_7 \ .
\end{align}
The dynamical system for this case is then given by
\begin{equation}\label{DynSystCurvAxion}
\begin{aligned}
    \frac{d x_1}{dN} &=-4+3x_1+2x_3-x_1x_3+x_1^2+4x_4-x_1x_4+6x_5^2 \\
    \frac{d x_2}{dN} &=8+m-4x_3+x_1x_2-2x_2x_3+4x_2-6x_4-2x_2x_4 \\
    \frac{d x_3}{dN} &= -8-m+8x_3-2x_3^2+6x_4-2x_3x_4 \\ 
    \frac{d x_4}{dN} &=-2x_3x_4-2x_4^2+4x_4-2x_4\\
    \frac{d x_5}{dN} &= -x_5+\frac{1}{2x_5}(4x_3+6x_4-8-m)-x_5(x_3+x_4)+\frac{1}{2}x_1x_5-\sqrt{\frac{3}{2}}x_6^2x_7 \\ 
    \frac{d x_6}{dN} &= \sqrt{\frac{3}{2}}x_5x_6x_7-x_6(x_3+x_4-2)+\frac{1}{2}x_1x_6\\ 
    \frac{d x_7}{dN} &= -\frac{1}{2}x_1x_7+\sqrt{6}x_5x_7^2(\Gamma-1) \ .  
\end{aligned}
\end{equation}
This system as well as containing the parameter $m$ contains also the parameter $\Gamma$ defined as
\begin{equation}\label{GammaV}
    \Gamma = \frac{V V_{\phi\phi}}{V_\phi^2} \ .
\end{equation}
In order to make this system autonomous along with $m$ also $\Gamma$ has to be a constant. This condition narrows the choices of the potential $V(\phi)$. For example, for exponential potentials ($V(\phi)=V_0e^{-\alpha \phi}$), which correspond to $\Gamma = 1$, it holds true (as this method was first found to render autonomous the dynamical system of an exponential potential theory \cite{copeland1998exponential}). It also holds true for potentials of the type $V(\phi)=V_0e^{-x_7 \phi}$, from the definition of $x_7$, if $\frac{\partial}{\partial \phi}\left(F'(R,\phi)\right)$ is negligible. In the case $\Gamma = 0$, we have a non-null solution for the potential to be $V(\phi) = V_0(1+\alpha \phi)$. \\
For constant $\Gamma$ the fixed points are found to be
\begin{align}\label{FixPointDM}
(x_1,x_2,x_3,x_4,x_5,x_6,x_7)=\phi_1^* &= (6,0,2,0,-i\sqrt{7},0,0) \\ 
    \phi_2^* &=\left(6,0,2,0,-i\sqrt{7},0,i\frac{\sqrt{\frac{3}{4}}}{\Gamma-1}\right) \\
    \phi_3^* &=(6,0,2,0,i\sqrt{7},0,0) \\ 
    \phi_4^* &=\left(6,0,2,0,i\sqrt{7},0,-i\frac{\sqrt{\frac{3}{4}}}{\Gamma-1}\right)\ .
\end{align}
The addition of the dark matter contribution has deeply changed the phase space scenario as can be seen by the expressions of the fixed points \eqref{FixPointDM} and by the stream plots in figure \ref{fig:phaseSpace}. We see that the fixed points depend on $\Gamma$ and seem to be divergent for $\Gamma = 1$. In this case, the dynamics of the $x_7$ variable becomes trivial since constant (and equal to zero), and therefore the fixed points are the same but in a phase space of type $(x_1,x_2,x_3,x_4,x_5,x_6)$. The presence of the imaginary unit in the fixed points (in particular in the variables $x_5$ and $x_7$) is not a problem since this will not appear in the physical quantities. This can be seen looking at the dynamical system in which we see that a choice of this type will lead to a overall imaginary coefficient in the equation for $\frac{d x_5}{d N}$ and $\frac{d x_7}{d N}$, and so this choice will modify the dynamic of the system (changing the signs of the terms) but actually without introducing imaginary units. We will see also at the end of this section that the during the computations of the various physical quantities the imaginary unit disappears from the final result. We see that all the fixed points are de Sitter attractors since $w_\text{eff}=-1$, as can be seen from Eq. \eqref{weffCurv}. We can perform a stability analysis for the case $m=0$ and $\Gamma=0$ analyzing the eigenvalues of the Jacobiam matrix of the dynamical system for each fixed point \eqref{FixPointDM}, as described in \cite{odintsov2017autonomous}, which for this case are found to be
\begin{align}
    \phi_1^* &\rightarrow (7,6,	6,	-3,	3,	-2,	0) \\
    \phi_2^* &\rightarrow \textstyle(7,	6,	6,	3,	-2,	\frac{3}{2},	0) \\
    \phi_3^* &\rightarrow (7,	6,	6,	-3,	3,	-2,	0) \\
    \phi_4^* &\rightarrow \textstyle(7,	6,	6,	3,	-2,	\frac{3}{2}	,0) \ . 
\end{align}
Since for each fixed point at least one eigenvalue has positive real part we conclude that there are no stable points for the dynamical system \eqref{DynSystCurvAxion}, as can be also deduced from the phase space plots in figure \ref{fig:phaseSpace}.

The dynamical variables \eqref{DynVarCurvAxion} can be rewritten in the holographic picture thanks to the holographic equation \eqref{holEq}. As in the non-flat case, we need to use the expression of the scale parameter \eqref{scaleParameterHol}. The corresponding dynamical variables are
\begin{equation}\label{DynVarCurvAxionHol}
    \begin{aligned}
  \hfill  &x_1 = -\frac{L_{IR} \, \dot{F'}}{c \, F'} \hspace{0.7cm} x_2 = -\frac{L_{IR}^2 \, F}{6 \, c^2 \,  F'} \hspace{0.7cm} x_3 = 2- \frac{\Dot{L}_{IR}}{c}+\frac{K L_{IR}^2}{c^2}e^{-2\int_{t_0}^t\frac{dt'}{L_{IR}}}\hfill \\
     &x_4=-\frac{K L_{IR}^2}{c^2}e^{-2\int_{t_0}^t\frac{dt'}{L_{IR}}} \hspace{0.3cm} x_5 = \frac{\kappa L_{IR}\dot{\phi}}{\sqrt{6}c\sqrt{F'}} \hspace{0.3cm} x_6 =\frac{\kappa L_{IR}\sqrt{V}}{\sqrt{3}c\sqrt{F'}} \hspace{0.3cm} x_7 = \sqrt{F'}\frac{V_\phi}{\kappa V}  \ .\hfill
\end{aligned}
\end{equation}
With this variables the system \eqref{DynSystCurvAxion} still holds for $m$ given in the non-flat vacuum $F(R)$ analysis \eqref{mCurved}. Compared to the standard method, the parameter $\Gamma$ is unchanged in the holographic frame and is still defined as \eqref{GammaV}.\\
We can analyze the expressions of the dynamical variables around the fixed points \eqref{FixPointDM}. As in the flat vacuum case we will study the condition $m\simeq0$, which corresponds to a cut-off of the type \eqref{cutOffmO}. The parameter $\Gamma$ will be treated as a constant different from one for the reasoning explained above. The fixed points can be summarized as 
\begin{align}
    \phi^*_{1,3} &= (6,0,2,0,\pm i \sqrt{7},0,0) \\
    \phi^*_{2,4} &= \left(6,0,2,0,\pm i \sqrt{7},0,\mp \sqrt{\frac{3}{4}}\frac{i}{\Gamma-1}\right) \ ,
\end{align}
from which we obtain the conditions on the dynamic variables as
\begin{equation}\label{CutOffFixedPoints}
    \begin{aligned}
  \hfill   &-\frac{L_{IR} \, \dot{F'}}{c \, F'} = 6\hspace{0.7cm} -\frac{L_{IR}^2 \, F}{6 \, c^2 \,  F'}=0 \hspace{0.7cm}  \frac{\Dot{L}_{IR}}{c}=\frac{K L_{IR}^2}{c^2}e^{-2\int_{t_0}^t\frac{dt'}{L_{IR}}}\hspace{1.cm}\hfill \\
  \hfill -&\frac{K L_{IR}^2}{c^2}e^{-2\int_{t_0}^t\frac{dt'}{L_{IR}}}=0
   \hspace{1.cm}  \frac{\kappa L_{IR}\dot{\phi}}{\sqrt{6}c\sqrt{F'}} =\pm i\sqrt{7} \hspace{1cm} \frac{ \kappa L_{IR}\sqrt{V}}{\sqrt{3}c\sqrt{F'}}=0 \hfill \\ 
   &\hspace{4.cm} \sqrt{F'}\frac{V_\phi}{\kappa V} =\left\{ \begin{aligned} &0 \\ \mp &\textstyle\sqrt{\frac{3}{4}}\frac{i}{\Gamma-1} \end{aligned}\right. \ , \hspace{2.cm}
\end{aligned}
\end{equation}
we can interpret the expressions of the fixed points as conditions on the cut-off and the other functions in the above system. Some of these conditions can be a good physically realistic definition for the infrared cut-off due to the link with the cosmology field equations expressed in terms of the autonomous dynamical system. As seen in the vacuum $F(R)$ case we need to be careful with the order in which we impose this conditions, as for example if we start imposing the condition on $x_2$ we would obtain a null $F(R)$, as explained under Eq. \eqref{FixedF(R)}. For example from the condition on the variable $x_5$ we can find an expression for $F(R)$. Using the static point condition on $x_5$
\begin{equation}
    \frac{\kappa L_{IR}\dot{\phi}}{\sqrt{6}c\sqrt{F'}} = \pm i\sqrt{7} \ ,
\end{equation}
and on $x_3 = 2$, we get the differential equation
\begin{equation}
    \frac{\partial F}{\partial R} = \mp \frac{2}{7}\frac{\kappa^2\dot{\phi}^2}{R}\ ,
\end{equation}
which can be integrated to obtain an expression for the $F(R,\phi)$ of the type
\begin{equation}
    F(R,\phi) = \Lambda(\phi) - \frac{2}{7}\kappa^2\dot{\phi}^2 \ln{R} \ . 
\end{equation}
For $x_7$ we have two cases. In both case we will need the condition on the variable $x_5$ rewritten in a manner that helps us to cancel the term $\sqrt{F'}$ which is 
\begin{equation}
    \sqrt{F'}=\pm\frac{\kappa L_{IR}\dot{\phi}}{i\sqrt{42}c} \ .
\end{equation}
If $x_7=0$ we find $\frac{\Dot{V}}{V}=0$, which shows that the potential has to be constant in time. More interesting is the case in which $x_7 = \mp \sqrt{\frac{3}{4}}\frac{i}{\Gamma-1}$, in this case the potential is constrained by the differential equation
\begin{equation}
    \frac{\dot{V}}{V} = 3\sqrt{\frac{7}{2}}\frac{1}{\Gamma-1}\frac{c}{L_{IR}} \ ,
\end{equation}
which can be integrated to give an explicit expression for the potential
\begin{equation}
    V(t) = V_0 e^{3\sqrt{\frac{7}{2}}\frac{1}{\Gamma-1}\left(H_0t-\frac{H_i}{2}t^2\right)} \ .
\end{equation}
As anticipated when we gave the expression of the fixed points the imaginary unit appearing in the variables $x_5$ and $x_7$ disappears when we compute physical quantities, which for example are the expression of $F(R)$ or the potential computed above. Furthermore as the vacuum case we should use the conditions on the other variables to fix conditions on the constant appearing in the final quantities.

\section{Conclusions}
In this work we tried to give an holographic description of the analysis of $F(R)$ models coupling axion dark matter and in particular the autonomous dynamical system method. After briefly introducing the $F(R)$ models and the holographic description of cosmology, we provided the analysis of phase space dynamics of the dynamical system given by the cosmological equations for different $F(R)$ models and rewritten this analysis in the holographic framework. The autonomous dynamical system formalism is known to give a powerful method to analyze unification models and their evolution, as well as the holographic framework, in which it is easy to obtain such unified models. In the holographic description the choice of the infrared cut-off is non-unique, and this freedom can allow us to find the best expression that simplifies the most the analytical computations. The general expression of this cut-off is given by the Nojiri-Odintsov cut-off \cite{nojiri2006unifying,nojiri2021different}, which is the cut-off that has been used in this work.
For these reasons, the conversion of the autonomous dynamical system analysis in the holographic language can give interesting conditions on the expression of the holographic cut-off. As seen in this paper, the fixed points in the phase space provide these conditions \eqref{CutOffFixedPoints} and we can see that each of these points corresponds to a $F(R)$ model and an infrared cut-off.\\
We performed the analysis of the autonomous dynamical system in the case of $F(R)$ models in a flat and non-flat FRW background, which was already known to provide promising results \cite{odintsov2017autonomous,odintsov2019effects}. Then we computed the case of $F(R)$ coupling axion dark matter in non-flat FRW, where the general relativity coupled axion dark matter case has been computed firstly in \cite{copeland1998exponential} and the generalization to $F(R)$ is straightforward but never given in literature as far as we know.

\section*{Acknowledgments}
This work is funded by MCIN/AEI/10.13039/501100011033 and FSE+, reference PRE2021-098098.

\phantomsection
\bibliographystyle{ieeetr}
\bibliography{Bibliography}

\end{document}